\documentclass[twocolumn,showpacs,preprintnumbers,amsmath,amssymb,prl,nofootinbib]{revtex4}

\usepackage{graphicx}
\usepackage{graphics}
\usepackage{dcolumn}
\usepackage{bm}
\usepackage{color}

\newcommand{\ttise}{1\textit{T}-TiSe$_2$}

\begin{document}

\preprint{APS/123-QED}
\bibliographystyle{prsty}

\title{Dramatic effective mass reduction driven by strong electronic correlations}

\author{C. Monney$^1$}
\author{E.F. Schwier$^1$}
\author{M.G. Garnier$^1$}
\author{C. Battaglia$^2$}
\author{N. Mariotti$^1$}
\author{C. Didiot$^1$}
\author{H. Cercellier$^3$}
\author{J. Marcus$^3$}
\author{H. Berger$^4$}
\author{A.N. Titov$^{5,6}$}
\author{H. Beck$^1$}
\author{P. Aebi$^1$}
\email{philipp.aebi@unifr.ch}

\affiliation{%
$^1$D\'epartement de Physique and Fribourg Center for Nanomaterials, Universit\'e de Fribourg, CH-1700 Fribourg, Switzerland\\
$^2$EPFL, Institute of Microengineering, Photovoltaics and Thin Film Electronics Laboratory, CH-2000 Neuch\^atel, Switzerland\\  
$^3$Institut N\'eel, CNRS-UJF, BP 166, 38042 Grenoble, France\\
$^4$EPFL, Institut de Physique de la Mati\`ere Condens\'ee, CH-1015 Lausanne, Switzerland\\
$^5$Institute of Metal Physics UrD RAS, Ekaterinburg, 620219, Russia\\
$^6$Institute of Metallurgy UrD RAS, Amundsen St. 101, Ekaterinburg, 620016, Russia
}%

\date{\today}

\begin{abstract}
We present angle-resolved photoemission experiments on \ttise\ at temperatures ranging from 13K to 288K. The data evidence a dramatic renormalization of the conduction band below 100K, whose origin can be explained with the exciton condensate phase model. The renormalization translates into a substantial effective mass reduction of the dominant charge carriers and can be directly related to the low temperature downturn of the resistivity of \ttise. This observation is in opposition to the common belief that strong interactions produce heavier quasiparticles through an increased effective mass.
\end{abstract}



\maketitle{}

Particularly stimulated by the discovery of high-temperature superconductivity, interest in strongly correlated electron systems never ceased to increase in the last decades. Strong correlations in e.g., heavy fermion systems involving $d$- and $f$-electrons usually produce weakly dispersive bands in their electronic structure, which give rise to quasiparticles with large effective masses \cite{StrongElCorrel}. Coupling of the electronic states to phonon modes is also well known as a possible mechanism for such an effect \cite{HengsbergerElPh}. In all these examples, strong correlations are commonly considered as responsible for important band renormalizations, leading to increased band effective masses. In the case of \ttise\ studied below, it will be shown that strong electronic correlations result in the opposite effect.
 
\begin{figure}
\centering
\includegraphics[width=7.6cm]{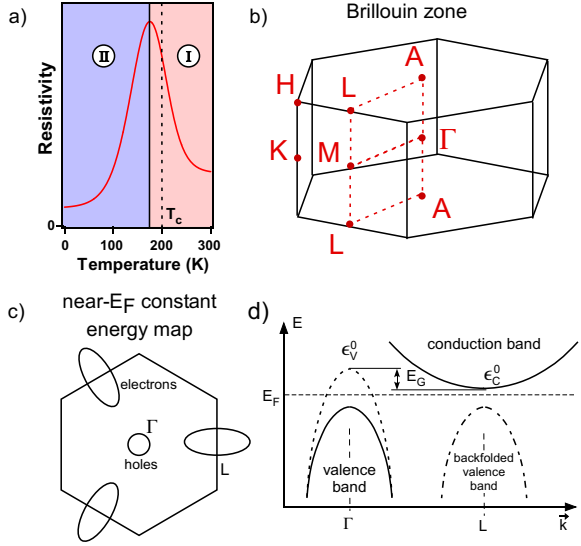}
\caption{\label{fig_1}
(Color online) (a) Schematic plot of the resistivity of \ttise, where two different regions are distinguished (adapted from \cite{DiSalvo}). (b) Brillouin zone of \ttise\ with its high symmetry points. (c) Schematized near-$E_F$ constant energy map of \ttise\ around $\Gamma$ and $L$, with electron pockets at $L$ produced by the conduction band. (d) At $\Gamma$, the gap opening due to the exciton condensation shifts the valence band far below $E_F$ and only the conduction band provides some electronic states near $E_F$.
}
\end{figure}

Among the transition metal dichalcogenides \cite{ClercTMDC}, \ttise\ is of particular interest. At the critical temperature of $T_c\simeq200$K, the system undergoes a phase transition into a charge density wave (CDW) phase accompanied by a periodic lattice distortion (PLD) \cite{DiSalvo}. Around the transition, it displays strongly anomalous transport properties. In particular, as the temperature decreases below 300K, the resistivity quickly increases and surprisingly, it falls back again to smaller values at lower temperatures (the resistivity is schematically reproduced in Fig. \ref{fig_1} (a)). The compound has also attracted strong interest since superconductivity has been discovered in the copper intercalated material, Cu$_x$TiSe$_2$ \cite{Morosan2006}, as well as in the pure material under pressure \cite{SiposSupra}. In the past years, different angle-resolved photoemission spectroscopy (ARPES) studies were carried out on \ttise \cite{PilloPRB,Kidd,Rossnagel,CercellierPRL}. They all evidenced two main contributions near the Fermi energy $E_F$, a Se-$4p$ derived band at the center of the Brillouin zone (BZ) ($\Gamma$ point on Fig. \ref{fig_1} (b)), identified as a valence band in what follows, and three (symmetry equivalent) Ti-$3d$ derived bands at the border of the BZ ($L$ point), identified as conduction bands (see Fig. \ref{fig_1} (c) and (d)). 

As the temperature decreases below $T_c$, an intense backfolded valence band appears at $L$ (dashed dotted line on Fig. \ref{fig_1} (d)), as direct evidence for the CDW. It is well known that, in the case of a CDW phase, the intensity in the backfolded bands is proportional to the strength of the new potential of competing periodicity \cite{Voit,ClercTaS2}. Based on the fact that in \ttise\ the backfolded valence band displays a high intensity despite the small atomic displacements ($\sim$0.08\AA) of the PLD \cite{DiSalvo}, we recently gave much support to the exciton condensate phase as the primary origin of the CDW phase in \ttise\ \cite{CercellierPRL}, which relies on strong electronic correlations, in agreement with the prediction of Wilson {\it et al.} \cite{WilsonComm,WilsonReview}.

Originally denominated the ``excitonic insulator phase'' \cite{Keldysh,Jerome}, the basic ingredients of this exotic phase are a valence and a conduction band, having a semimetallic or semiconducting configuration. Then, bound states of holes and electrons, called excitons, can condense at low temperature into a macroscopic state, provided the gap is small and the screening of the Coulomb interaction is weak. For \ttise, this purely electronic effect naturally generates the CDW \cite{MonneyPRB}. Since an exciton is a neutral quasiparticle, condensation of such entities removes charge carriers from the system. In fact, this effect already occurs above $T_c$ due to strong electron-hole fluctuations \cite{MonneyTScan}, so that the resistivity starts to increase above $T_c$ (region I in Fig. \ref{fig_1} (a)). Exciton condensation manifests itself in the band structure by a gap opening below the conduction band, which shifts the valence band away from $E_F$ \cite{MonneyPRB}. However, all along the transition, the conduction band remains close to $E_F$, providing occupied states to transport so that the system never really becomes insulating (see Fig. \ref{fig_1} (d)).  

In this letter, we present temperature dependent ARPES measurements of \ttise. For the first time, a dramatic renormalization of the conduction band at low temperature is evidenced. Its origin is explained using the exciton condensate phase model. It is related to a strong reduction of the electron effective mass and delivers an essential component to understand the resistivity downturn at low temperature (region II in Fig. \ref{fig_1} (a)).

\begin{figure}
\centering
\includegraphics[width=8.1cm]{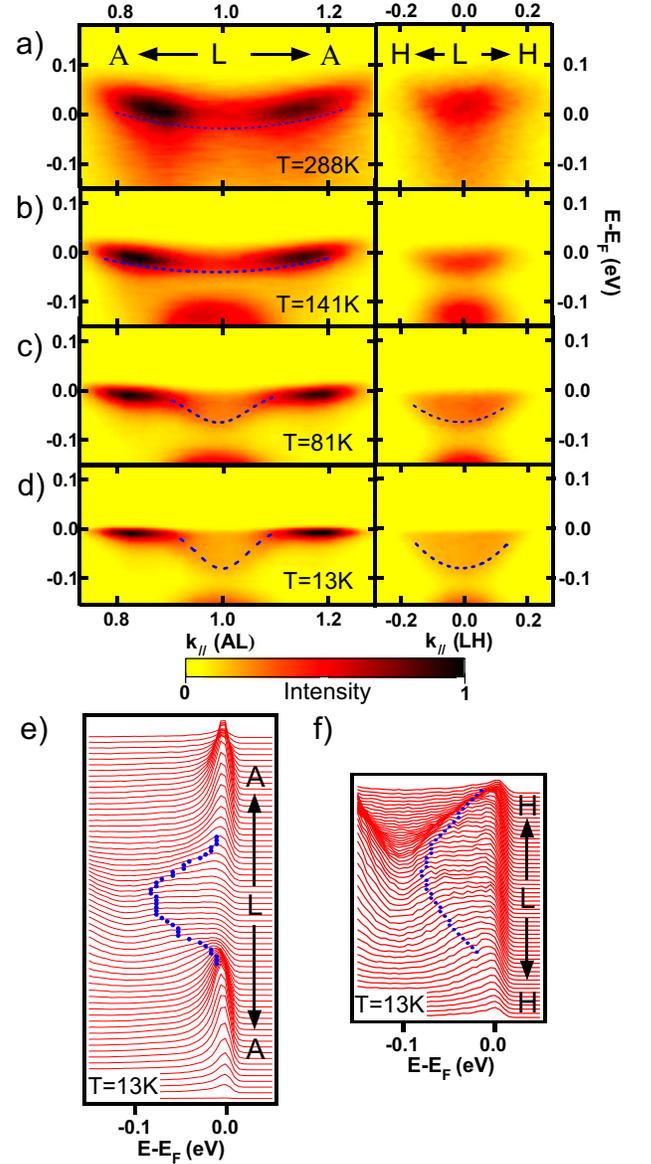}
\caption{\label{fig_2}
(Color online) Photoemission intensity maps (dark colours represent high intensity) of the electron pocket around $L$. (a), (b), (c), (d) Measurements along $AL$ (left panels) and $LH$ (right panels) at 288K, 141K, 81K and 13K respectively. (e), (f) Energy distribution curves of the photoemission intensity maps measured at 13K around $L$ and along $AL$ and $LH$, respectively. The blue dots indicate the position of the high binding energy edges (see text).
}
\end{figure}

The photoemission intensity maps presented here were recorded using monochromatized, linearly $p$-polarized HeI$\alpha$ radiation at 21.2 eV and using an upgraded Scienta SES-200 spectrometer with an overall energy resolution better than 10 meV. A liquid helium cooled manipulator having an angular resolution of $~0.1^\circ$ was used, with a temperature stability $<5$K. \ttise\ samples were cleaved \textit{in-situ}, in a base pressure in the low $10^{-11}$ mbar, ensuring a high longevity of the sample surface. Photoemission spectra were recorded from 13K to 288K. At the end of the measurements, the sample was cooled to 13K and comparable spectra were recorded again, confirming its stability. Reference spectra of polycrystalline gold evaporated on the same sampleholder were recorded for determining $E_F$. At the excitation energy of 21.2 eV, at the border of the BZ, initial states close to the $L$ point are probed (see the BZ depicted in Fig. \ref{fig_1}(b)). Therefore, to simplify, the notation $L$ will be chosen for these states throughout this article.

For \ttise, below the critical temperature, the main feature in the neighbourhood of $E_F$ is the conduction band at $L$, since the valence band at $\Gamma$ is shifted to higher binding energies (see Fig. \ref{fig_1} (d)) \cite{CercellierPRL}. Fig. \ref{fig_2} presents photoemission intensity maps of this electron pocket taken at four different temperatures, along the high symmetry directions $AL$ and $LH$. Binding energy versus momentum carpets highlight the behaviour of the dispersion of the conduction band at 288K, 141K, 81K and 13K in Fig. \ref{fig_2} (a), (b), (c) and (d), respectively. At 288K, the conduction band follows a clear wide parabolic dispersion along $AL$ (left panel), with a non-trivial spectral weight distribution that can be understood within the exciton condensate phase model extended to the strong electron-hole fluctuation regime \cite{MonneyPRB}. 
As the temperature decreases to 141K (Fig. \ref{fig_2} (b)), the conduction band gets flatter along $AL$ (left panel). The top of the backfolded valence band is also clearly visible below a binding energy of $-0.1$ eV. Below 100K, a dramatic renormalization of the conduction band is visible, as seen in the photoemission intensity maps measured at 81K and 13K (Fig. \ref{fig_2} (c) and (d)). Along $AL$ (left panels), it has no more a simple parabolic shape and divides into different parts. Its branches closer to $E_F$ are getting flatter, while it displays a pronounced parabolic dispersion in the neighbourhood of its minimum. The $LH$ carpets (right panels) show a homogeneous spectral weight distribution.

In order to get the approximate behaviour of the quasiparticle dispersion within the temperature dependent spectral weight distribution close to the Fermi-Dirac cutoff, we perform parabolic fits of the high binding energy edge of the quasiparticle dispersion (defined here as the energy position at which, for each energy distribution curve (EDC), the intensity reaches $\sim 90\%$ of the maximum intensity of the dispersion). This procedure determines the curvature of the dispersions without being strongly affected by intensity variations and by the Fermi-Dirac cutoff varying with temperature \cite{footnote}.
These fits allow us to estimate the effective mass of the conduction band near its minimum (at $L$). We show in Fig. \ref{fig_2} (e) and (f) the high binding energy edges (blue dots) obtained for the photoemission intensity maps measured at 13K (Fig. \ref{fig_2} (d)) around $L$ and along $AL$ and $LH$, respectively, and the corresponding EDCs. 
Table \ref{tab_1} summarizes these results. This renormalization is so strong that $m_L$ is reduced by about a factor 10-15 from 288K to 13K.
\begin{table}[ht]
\caption{Renormalized effective mass of the conduction band (in units of the bare electron mass) along the long axis $\tilde{m}_L$ (along AL) and the short axis $\tilde{m}_S$ (along LH) of its elliptic Fermi surface, as a function of temperature.  
}
\begin{tabular}{p{1cm}p{1cm}p{1cm}p{1cm}p{1cm}}
\hline\hline
 & $13$K & $81$K & $141$K & $288$K \\ 
\hline
$\tilde{m}_L$ & 0.4(2) & 0.5(2) & 6.6(7) & 6.2(7) \\
$\tilde{m}_S$ & 0.5(2) & 0.8(2) & \footnotemark[1] & \footnotemark[1] \\
\hline\hline
\end{tabular}
\footnotetext[1]{At these temperatures, the procedure to extract $\tilde{m}_s$ becomes inaccurate due to the proximity of the Fermi cutoff.}
\label{tab_1}
\end{table}
\begin{figure}
\centering
\includegraphics[width=8.cm]{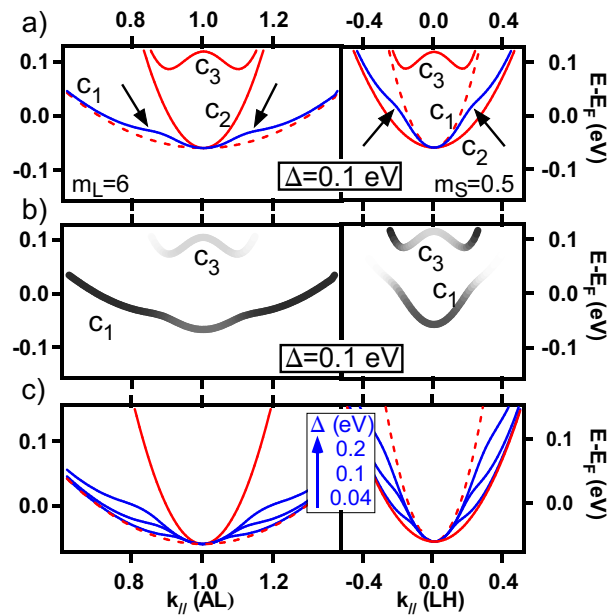}
\caption{\label{fig_3} 
(Color online) Renormalized band dispersions calculated within the exciton condensate phase model, around $L$, for $\Delta=0.1$ eV (a) without the spectral weights and (b) with the spectral weights. The effective masses of the conduction band are fixed to $m_L=6$ and to $m_S=0.5$. Dashed lines represent bands calculated with an order parameter $\Delta=0$ eV.  The blue line represents the main conduction band $c_1$. (c) Renormalization of the main conduction band as a function of the order parameter, for $\Delta=0.04,0.1,0.2$ eV. The conduction band $c_3$ is omitted for clarity.
}
\end{figure}

The origin of the considerable renormalization of the conduction band can be understood in the framework of the exciton condensate phase model. In this model, the CDW phase is naturally realized, when the order parameter $\Delta$ of the exciton condensate phase acquires non-zero values \cite{MonneyPRB}. The electron-hole interaction, responsible for the formation of excitons, couples the three conduction bands to the valence band, so that in the CDW phase, many backfolded bands appear at $L$, giving rise to a complicated band structure. Fig. \ref{fig_3} (a) depicts the situation near $E_F$ for the renormalized dispersions without the spectral weights. Continuous lines represent the main conduction band (blue) $c_1$ which cohabits with two symmetry equivalent backfolded conduction bands (red) $c_2$ and $c_3$, all calculated with an order parameter of $\Delta=0.1$ eV. The dashed (red) line represents the original ($\Delta=0$ eV) conduction band. We see that, for $\Delta=0.1$ eV and along AL, the main conduction band (blue) $c_1$ is bent near its minimum (arrows in Fig. \ref{fig_3} (a) (left)) towards the backfolded conduction band (red) $c_2$, giving rise to the mass renormalization observed in our ARPES data. It must be emphasized that this effect does not occur in a model where only \textit{one} conduction band is coupled to the valence band \cite{HalperinSSP}.
Fig. \ref{fig_3} (b) depicts the same situation, but now the spectral weights are superimposed on the dispersions in a grayscale coding. In the CDW phase, the spectral weight is transferred from the original band $c_1$ (dashed red curve in Fig. \ref{fig_3} (a)) to the backfolded band $c_3$ (and also to the backfolded valence band which is not shown in Fig. \ref{fig_3}, as it appears at higher binding energies), but mainly remains on the main conduction band $c_1$. The conduction band $c_2$ carries a negligible spectral weight and is thus not visible. Comparing Fig. \ref{fig_3} (a) and (b), we see that the mass renormalization affects mainly the main conduction band $c_1$ along $AL$, since along $LH$ this happens above $E_F$ (arrows in Fig. \ref{fig_3} (a) (right)). To better understand this effect, we plot in Fig. \ref{fig_3} (c) the main conduction band $c_1$ (blue) for different values of the order parameter $\Delta$. The lowest lying curve has been calculated with $\Delta=0.04$ eV, the intermediary one with $\Delta=0.1$ eV and the highest lying one with $\Delta=0.2$ eV. Along $AL$ (left panel), these calculations clearly show that, while the $\vec{k}$ range over which this band is concerned by the mass renormalization increases with $\Delta$, the reduced effective mass near $L$ is always the same, $\tilde{m}_L\simeq 1.6$ (obtained with parabolic fits close enough to its minimum). Along $LH$ (right panel), the situation is different, since the main conduction band (blue) always follows its original dispersion (dashed red) near $L$ and undergoes a renormalization (with increased effective mass) only away from $L$, where mainly empty states (invisible to ARPES) are affected. 

In our calculations, we have also seen that the ratio of the bare effective masses $m_L/m_S$ (i.e. without excitonic effects, at $\Delta=0$ eV) plays an essential role in this mass renormalization. The larger the anisotropy of the electron pocket, the stronger the renormalization effect and thus the lower the effective mass $\tilde{m}_L$. In Fig. \ref{fig_3}, we used a ratio of $m_L/m_S=12$, which is close to what is observed ($m_L/m_S\simeq 10-15$), and is already high enough to produce a substantial renormalization.

These results are in good agreement with our experimental observations, except for the calculated value of $\tilde{m}_L\simeq 1.6$ which underestimates the strength of the renormalization inferred from our ARPES data. We emphasize here that the exciton condensate phase model, which gives rise to the observed CDW, relies only on electronic degrees of freedom \cite{CercellierPRL,MonneyPRB}. However, the PLD, which is concomitant to the CDW and which is not considered in our model, may enhance the anistropy of the electron pocket, thus giving rise to a larger renormalization and reconciling the theoretical value of $\tilde{m}_L$ with the experimental one.

The substantial renormalization of the conduction band unveiled in this work must clearly have a strong influence on the resistivity of \ttise. Table \ref{tab_1}, which quantifies approximatively this effect, shows that below 100K the effective mass of the conduction band is reduced by a factor $\sim 10-15$. Fig. \ref{fig_1} (a) recalls the behaviour of this resistivity, where two regions are distinguished. As the temperature decreases below 300K, the resistivity quickly increases (region I). In the framework of the exciton condensate phase, a gap, already present above $T_c$ due to electron-hole fluctuations, increases strongly upon exciton formation and spectral weight is removed from the conduction band at the same time \cite{MonneyPRB,MonneyTScan}. Therefore, the number of charge carriers near $E_F$ available for transport is reduced and the resistivity increases. Below $T_c$, although the mechanism at work in region I still strengthens, different compensating effects enter into play. Below 100K, the dramatic effective mass renormalization reduces the resistivity (region II) and the scattering rate also decreases below $T_c$, emphasizing this effect \cite{LiOptics}. 

In conclusion, we have evidenced with photoemission for the first time a dramatic renormalization in the band structure of \ttise\ at $T<100$K. In this compound, the conduction band represents the main contribution near $E_F$. Below 100K, its effective mass decreases by a factor 10-15 with respect to its room temperature value. In the framework of the exciton condensate phase model, we are able to reproduce qualitatively this effect, which means that strong electron-electron interactions are responsible for it. Therefore, in opposition to the common belief, strong correlations in \ttise\ give rise to a substantial effective mass \textit{reduction} of the dominant charge carriers and, as a consequence, to a strong conductivity enhancement. This is, to our knowledge, the first observation of such a phenomenon.

\begin{acknowledgments}
We thank F. Baumberger and L. Forr\`o for valuable discussions.
We wish to acknowledge the support of our mechanical workshop and electronic engineering team. This project was supported by the Fonds National Suisse pour la Recherche Scientifique through Div. II and the Swiss 
National Center of Competence in Research MaNEP. 
\end{acknowledgments}

\end{document}